**Medium – long term earthquake prediction by the use of the oscillating electric field (T = 365 days) generated due to Earth's orbit around the Sun and due to its consequent oscillating lithospheric deformation.**

Thanassoulas[1], C., Klentos[2], V., Tsailas[3]., Verveniotis, G[4]., Zymaris, N. [5]

1. Retired from the Institute for Geology and Mineral Exploration (IGME), Geophysical Department, Athens, Greece.
   e-mail: thandin@otenet.gr - URL: www.earthquakeprediction.gr

2. Athens Water Supply & Sewerage Company (EYDAP),
   e-mail: klenvas@mycosmos.gr - URL: www.earthquakeprediction.gr

3. Institute for Geology and Mineral Exploration (IGME), Geophysical Department, Athens, Greece.

4. Ass. Director, Physics Teacher at 2[nd] Senior High School of Pyrgos, Greece.
   e-mail: gver36@otenet.gr - URL: www.earthquakeprediction.gr

5. Retired, Electronic Engineer.

**Abstract.**

We study the Earth's electric field monitored at PYR (Greece) monitoring site, for a period of more than six years (May 23[rd], 2003 to September 7[th], 2009). It is compared, in particular its oscillating component of T = 365 days, with the Perihelion – Aphelion dates of the Earth's orbit around the Sun, with the same component of the Earth's magnetic field, with the corresponding same period tidal oscillation and with the occurred large EQs of the same period of time. The obtained results suggest that the oscillating electric field component is generated by large scale piezoelectricity triggered by the Earth's shape – lithospheric oscillating deformation. The driving mechanism (yearly tidal variation) precedes the Aphelion – Perihelion dates for a month complying with the corresponding tidal friction behavior of the Earth's shape deformation. The Earth's oscillating electric field peaks coincide with the Perihelion – Aphelion dates while the triggered large EQs are clustered very close to the Perihelion – Aphelion dates. Moreover it is shown that the observed Earth's oscillating electric field is not related to or induced by the corresponding Earth's magnetic field. In conclusion, the Earth's oscillating electric field character could be used as medium to long term electric seismic precursor of large EQs.

**Keywords:** tidal waves, Earth deformation, preseismic electric signals, magnetic field, Perihelion, Aphelion, earthquake prediction.

## 1. Introduction.

The generation of earthquake precursory electric field, observed on ground surface, has been reported by many researchers in the past. Mechanisms that justify its generation have been presented in summary by Thanassoulas (2007) and in detail at the corresponding references therein. Fewer researchers have rejected the presence of earthquake precursory electric signals on the ground that these are generated by industrial, anthropogenic, ionospherically induced noise or even more by noise generated inside the data acquisition system (hardware malfunction). Characteristic are the cases presented by Pham et al. (2001) who attributed the origin of seismic electric signals (SES) of VAN group to the leakage of electric and phone networks and Pham et al. (2002) who suggested that the SES, recorded, by the VAN group at Lamia area, Greece, were of anthropogenic origin.

In real nature, all these electric fields / signals coexist in an actual recording of the Earth's electric field recorded on its ground surface. Therefore, the problem that is faced is: to find a way to discriminate between "noise" of any kind and (if any) "earthquake precursory electric signals", while it is important from a physical point of view to adopt a physical model that accounts for the generation of the observed earthquake precursory electric signals. Such methods and models were presented by Thanassoulas (2007).

In this work an analysis is applied on the Earth's electric field recorded for a rather long period of time (23[rd] May, 2003 to 7[th] September, 2009) by PYR (see figure 1) monitoring site, in Greece. The aim of this work is to study the corresponding electric field for its very long variations and if possible to suggest an appropriate model that can account for it. Moreover, a correlation will be attempted with the large EQs (Ms>6.5R) which occurred during the study period.

## 2. Data presentation and analysis.

Long wavelength electric fields are possible to identify only when longer periods of recordings are made. For the case of this work, the recorded electric field for a period of more than six years that is from May 23[rd], 2003 to September 7[th], 2009 (20030523 – 20090907) at PYR monitoring site in Greece will be studied. During this period of time eight (8) large EQs with magnitude over 6.5R (Ms > 6.5R) occurred. These EQs are tabulated in the following table (1) while their magnitude is presented as ML. It is pointed out that the ML value is converted to the corresponding Ms by adding 0.5 (Ms= ML+0.5). The date is presented in yyyymmddhhmm format.

TABLE – 1

| No. | Date | Lat. | Lon. | Z | ML |
|---|---|---|---|---|---|
| 1 | 200403170520 | 34.46 | 23.26 | 24 | 6.0 |
| 2 | 200601081134 | 36.21 | 23.41 | 69 | 6.4 |
| 3 | 200801060514 | 37.11 | 22.78 | 86 | 6.1 |
| 4 | 200802141009 | 36.50 | 21.78 | 41 | 6.2 |
| 5 | 200802141208 | 36.22 | 21.75 | 38 | 6.1 |
| 6 | 200802201827 | 36.18 | 21.72 | 25 | 6.0 |
| 7 | 200806081225 | 37.98 | 21.51 | 25 | 6.5 |
| 8 | 200807150326 | 35.85 | 27.92 | 56 | 6.2 |

The EQs location is shown in the following map of Greece of figure (1) as numbered blue circles. The background of the map presents in brown colors the seismic potential of Greece calculated for the year 2000, while the gray thick lines represent the fracturing of the lithosphere as it is deduced by the study of the gravity field of Greece (Thanassoulas, 2007). In the same map the location of the PYR monitoring site is shown with red lettering (aside EQ No. 7).

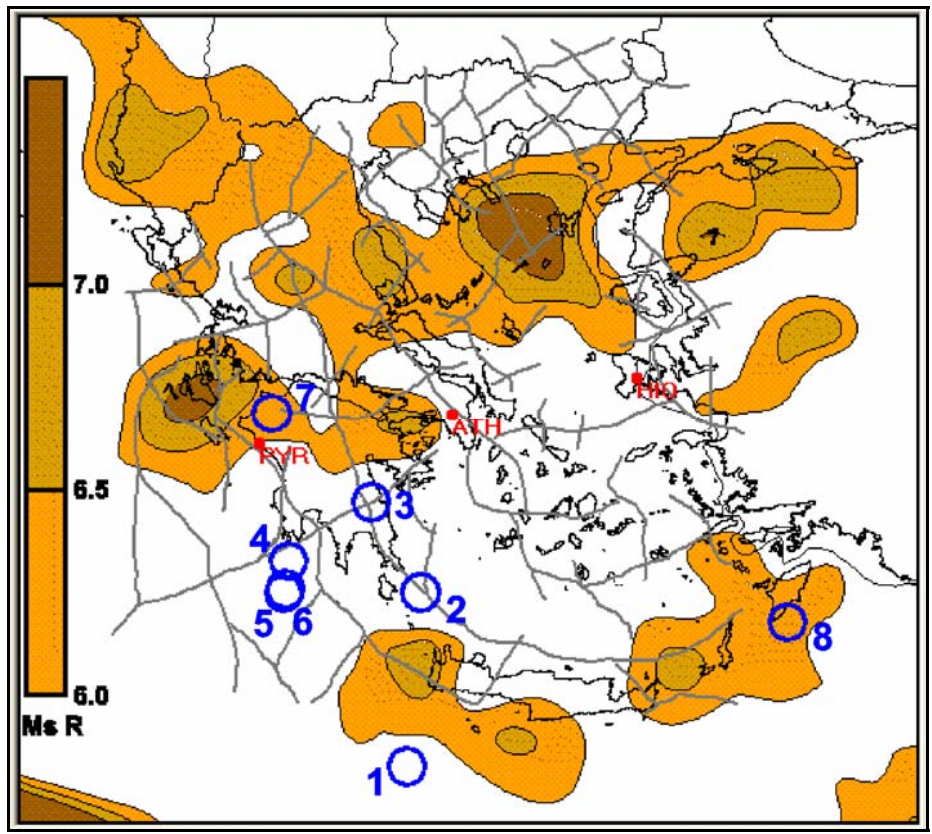

Fig. 1. Location of EQs (blue circles) on top of spatial distribution of seismic potential in Greece (brown colors) and the lithospheric fracturing (thick gray lines).

Although the entire recording network consists of three (ATH, PYR, HIO) monitoring sites, only PYR was selected for this study for the following reasons: HIO monitoring site is in operation for only three years, ATH monitoring site operates with a rather short receiving dipole (20m) compared to the HIO dipole of 200m, while PYR operates with a dipole of 160m and continuously for more than six years. Technical details for the monitoring sites have been presented by Thanassoulas (2007).

The recorded electric field by PYR monitoring site, for the study period, is presented in the following figure (2). Both E-W and N-S components of the recorded electric field are presented (black line). A three days sample recording is shown in the upper graph while in the lower one the entire study period, along with the large EQs (red bars) that occurred in the same period of time, are shown. The numbering below the graph of figure (2, lower) corresponds to the numbering of figure (1).

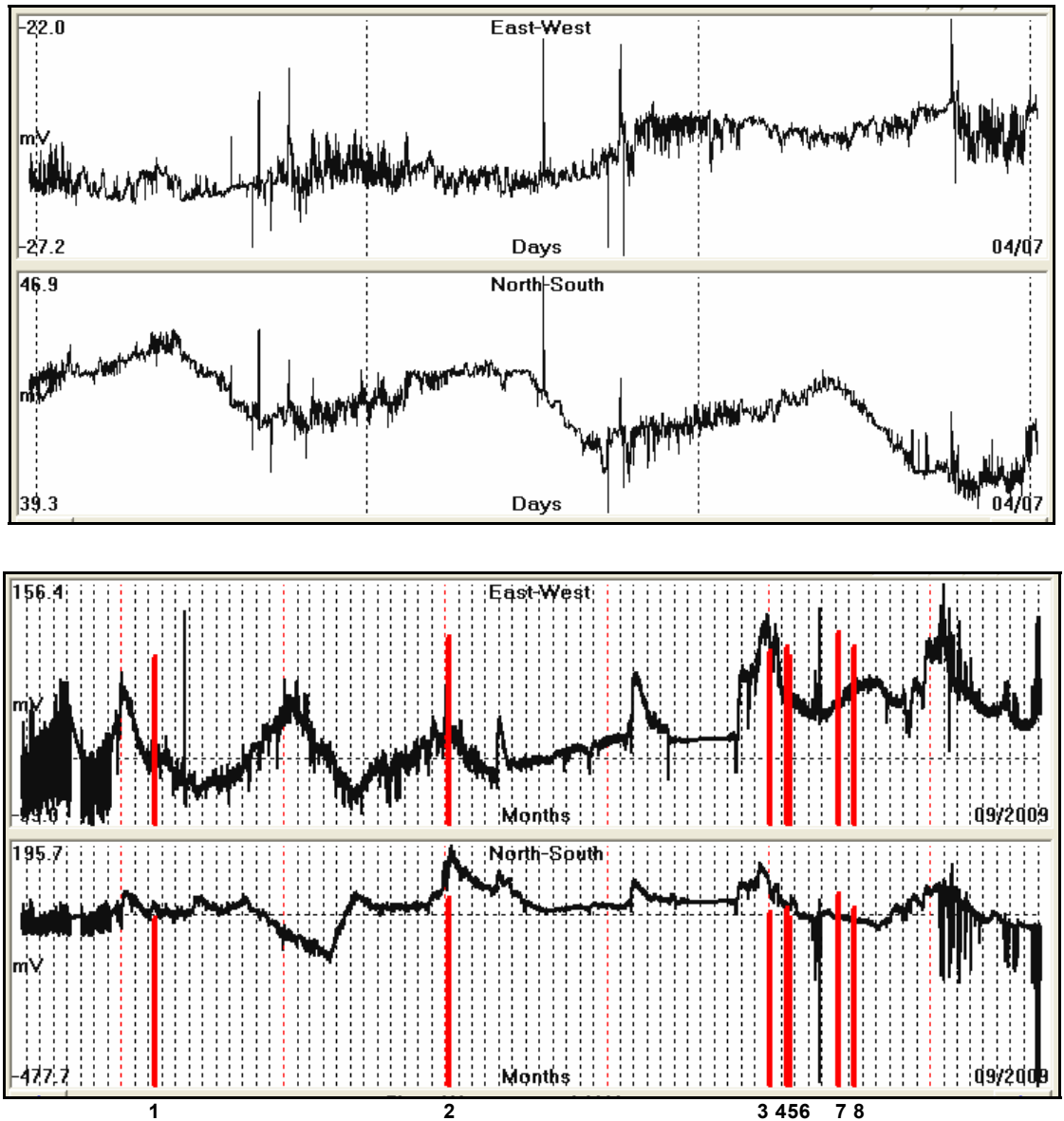

Fig. 2. PYR raw data (black line) of the Earth's electric field recorded for the period 20030523 – 20090907 along with the occurred large EQs (Ms >6.5R. Numbering corresponds to TABLE – 1. The upper graph shows a three days sample recording.

At a first glance both (E-W, N-S) recordings are characterized by some peculiar long wavelength noise. The latter is presented as characteristic peaks of the electric field of considerable amplitude compared to the short wavelength existing noise. The long wavelength noise is more evident in the E-W component than the same noise presented in the N-S one. The origin of this specific noise is at the present unknown. Moreover, the E-W component presents an increase (linear drift) in its amplitude of some decades of mV during the study period, while the N-S component is more stable. The observed linear drift could be initially attributed to either electrodes polarization drift or to corresponding hardware recording channel drift. In either case, this drift is ignored since we are interested in to the wider frequency content of the electric field and not to its DC components.

A different way to view the recorded raw data can be utilized by obtaining its corresponding frequency spectrum (Bath 1974, Kulhanek 1976, Claerbout 1976). The results of the transformation of the time-domain data into its frequency domain is presented in the following figure (3).

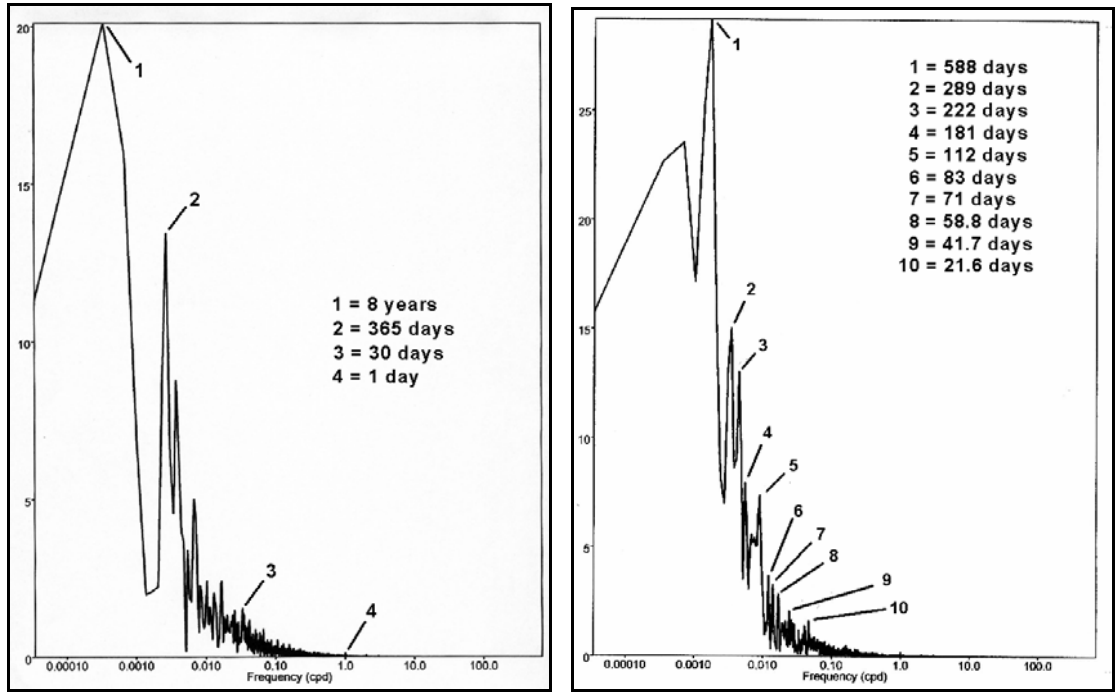

**Fig. 3.** PYR frequency spectrum of raw E-W (left) and N-S (right) data recorded at PYR monitoring site and distinct frequency components. The horizontal scale represents cycles per day (cpd). Numbering in the graph indicates (see inner table) specific periods of the various identified components of the analyzed electric field.

The left part of figure (3) represents the frequency spectrum of the E-W component. Obviously, the eight (8) years component (no. 1) is attributed to a fictitious result due to the linear drift since the data span for a period of time less (6 years) than the calculated period of eight (8) years. The next remarkable in amplitude value component corresponds to a period of T = 365 days, while just visible is the component of period of 30 days. The right part of figure (3) represents the frequency spectrum of the N-S component. In this graph the first thing to notice is the total absence of the 365 days period component while the larger periods have increased amplitude. The latter is rather justified by the absence of the recorded electric field linear drift.

The inspection of figures (2) and (3) suggests the presence in the recordings of some considerable noise with periods less than some days. Therefore, aiming to study very long period electric signals, we applied a low-pass filtering procedure on the raw data so that the frequency content of its spectrum with periods less than (14) days was totally rejected. The result of this operation is presented in the following figure (4).

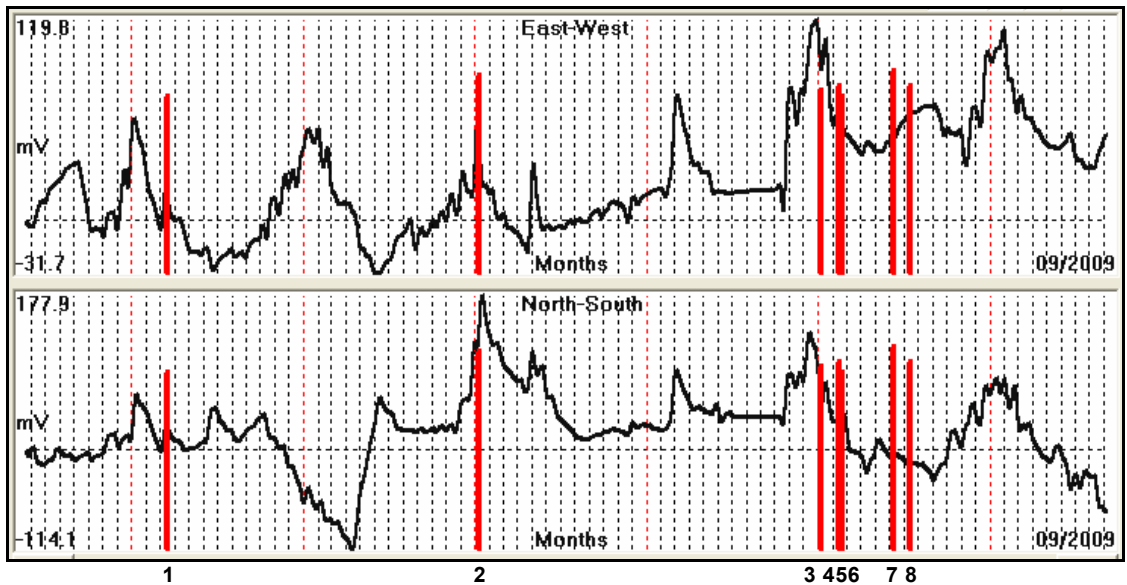

**Fig. 4.** PYR low-pass (T>14 days) filtered data plus large (red bars, Ms> 6.5R) EQs. Numbering corresponds to TABLE – 1.

What is evident from figure (4) is the presence of an oscillating component of the electric field with period of almost 12 months. Consequently, a band-pass filtering (with center period of T = 365 days) is applied on the raw data so that the oscillating component of T = 365 days is isolated. This operation is presented in the following figure (5)

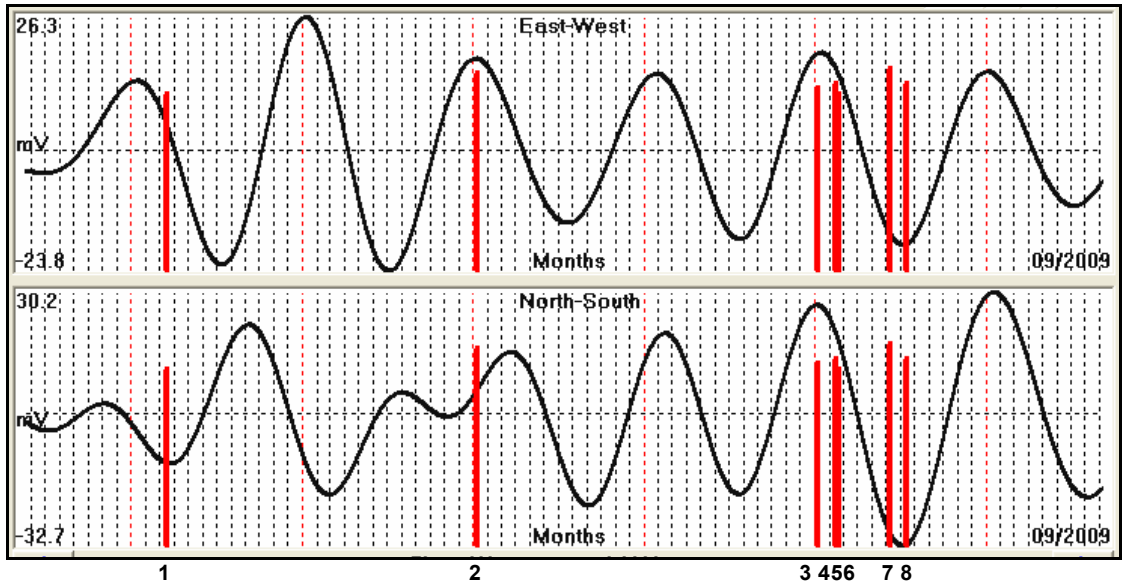

**Fig. 5.** PYR band-pass (T = 365 days) filtered data plus large (red bars, Ms> 6.5R) EQs. Numbering corresponds to TABLE – 1.

What is interesting in this drawing is firstly the fact that the E-W component presents a rather stable amplitude, while the N-S component increases its amplitude only during the last two (2) years when more large EQs occurred in a rather short time period. Moreover, during the first three years, apart from the amplitude difference between the E-W and N-S components, there is a considerable phase shift between them which diminishes during the last two years when the seismic activity is more intense (six large EQs within six months!) and clustered very close to the peaks of the oscillating electric field while at the same time both components become in phase.

The signal to noise ratio (S/N) in decibels above noise level (at 1 day period) of the T = 365 days of the E-W component has been calculated as S/N = +22.3 db.

Consequently, a question arises as what is the driving mechanism of this oscillating field. Generally, the oscillating electric field could be triggered by either "external" or "internal", as far as it concerns the solid Earth, mechanisms. The term "external" refers to any mechanism above ground surface such as i.e. the ionospheric induction currents in the earth. It is well known that the original magnetic field of the Earth is modulated by the ionospheric currents thus, in turn it could induce similar currents in the ground. The term "internal" refers to any physical mechanism that takes place in the ground, which can produce currents in it by modifying its regional physical parameters.

For the specific case of the oscillating Earth's electric field of T = 365 days we must obviously exclude the case of anthropogenic and industrially generated noise. Therefore, the Earth's oscillating field could be induced in the ground either "externally" by changes of the Earth's magnetic field generated by random ionospheric currents or "internally" by very slow and long period processes which take place in the interior of the Earth. Such an "internal" mechanism is the Earth's lithospheric deformation due to tidal waves.

Therefore, the corresponding Earth's total magnetic field and the tidal variations will be analyzed for the entire study period as far as it concerns their frequency component of T=365 days content. The magnetic data were provided by the Magnetic Observatory of Penteli (MOP) operated by the Geophysical Department of the Institute of Geology and Mineral Exploration (IGME) located in Athens Greece.

### 2.1. Magnetic total field data analysis.

The total magnetic field of the Earth and its ionospheric fluctuations is firstly hypothesized as the generating (by induction currents) mechanism of the oscillation of the Earth's electric field with T = 365 days. The latter is presented a: as a three (3) days sample recording (upper graph) and for the entire study period (lower graph) in the following figure (6).

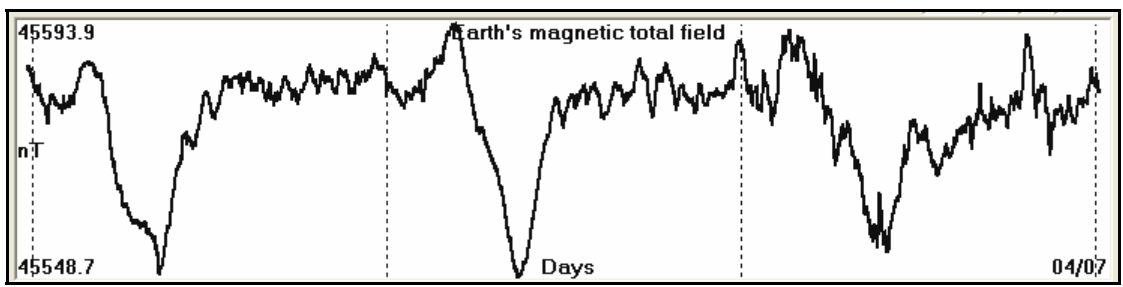

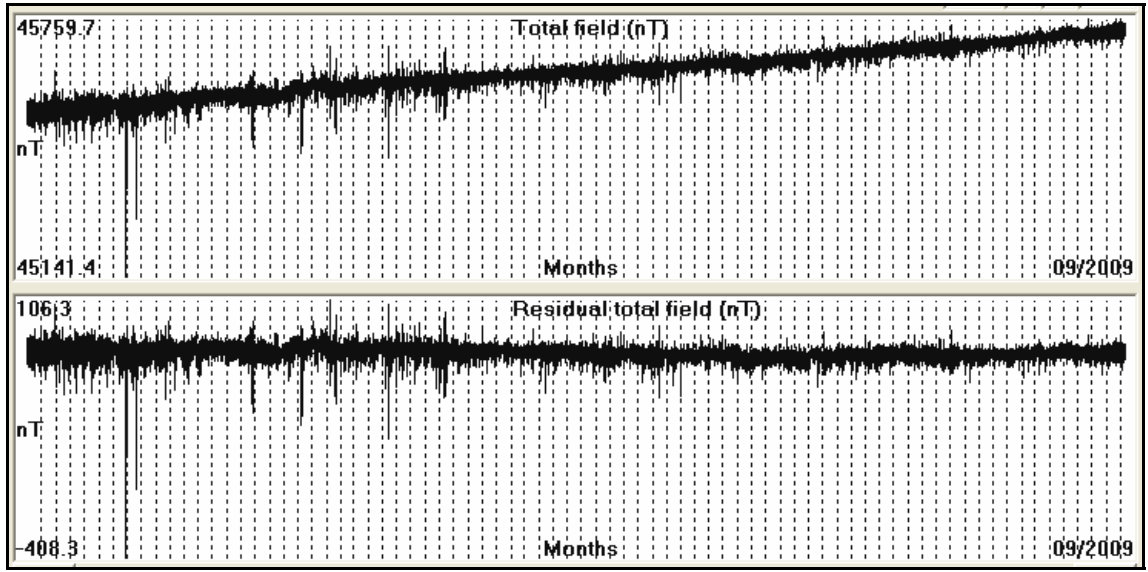

**Fig. 6.** Magnetic total field raw recording. A three days sample recording is shown in the upper graph while raw and detrended data for the entire study period are presented in the lower graph.

The upper graph of figure (6 lower) represents the raw total magnetic field as it was recorded by MOP, while the lower graph shows the same data after linear detrending and subtraction of a constant magnetic level of 45600nT. Furthermore, a frequency spectral analysis is applied on the time domain data in order to get their frequency content. The results of this operation are presented in the following figure (7).

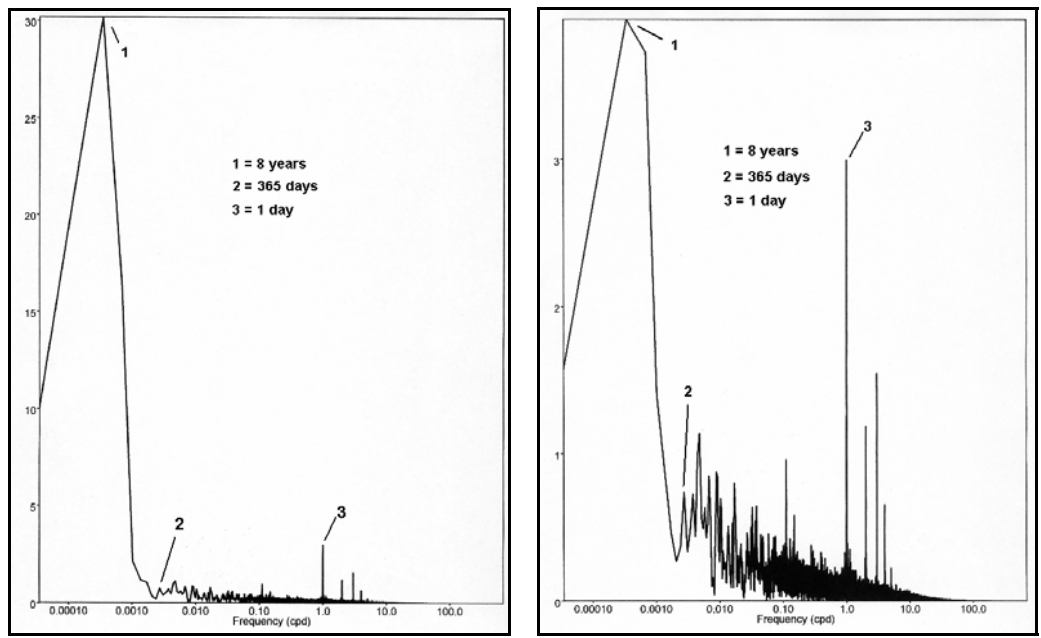

**Fig. 7.** Frequency spectral analysis of the total magnetic field of the study period. Raw magnetic data (left) and detrented data (right).

In both graphs the T = 8 years component is the dominant one, although with smaller amplitude in the detrended data graph. What is very interesting is the fact that the magnetic component with T = 365 days is very small even compared to the component with T = 24 hours. The latter is better visualized in the time domain by comparing the detrended data to the specific oscillating (T = 365 days) component (obtained by band-pass filtering) by a simultaneous presentation in a graph with same vertical scale for both data sets. This is shown in the following figure (8).

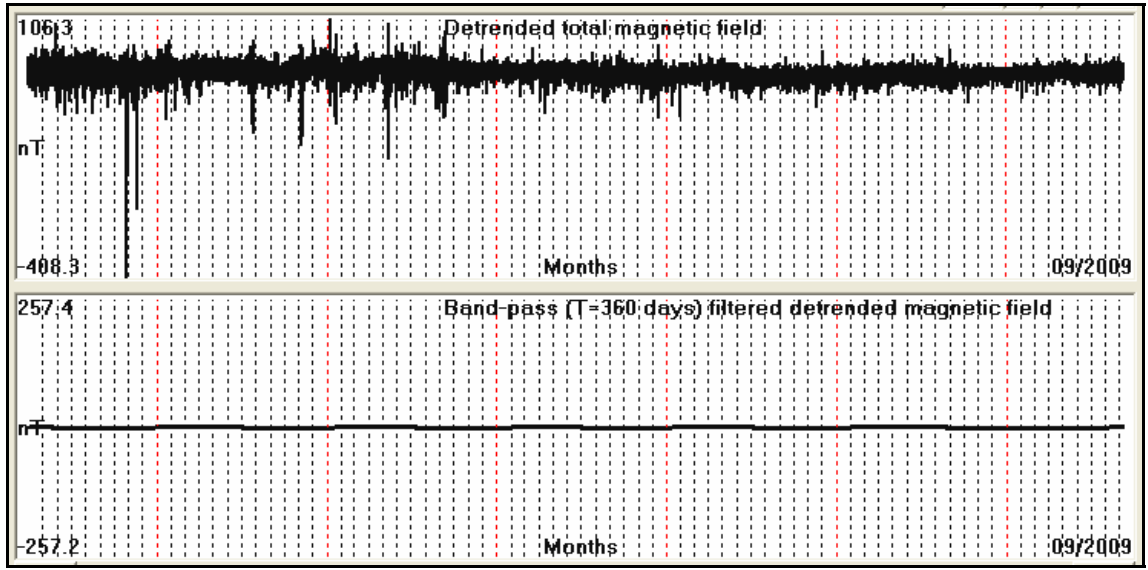

**Figure 8.** Detrended magnetic raw data (upper graph) and its oscillating component (band-pass filtered with T = 365 days).

**Additionally, the same data sets are presented in figure (9) in normalized vertical amplitude to the maximum graph range.**

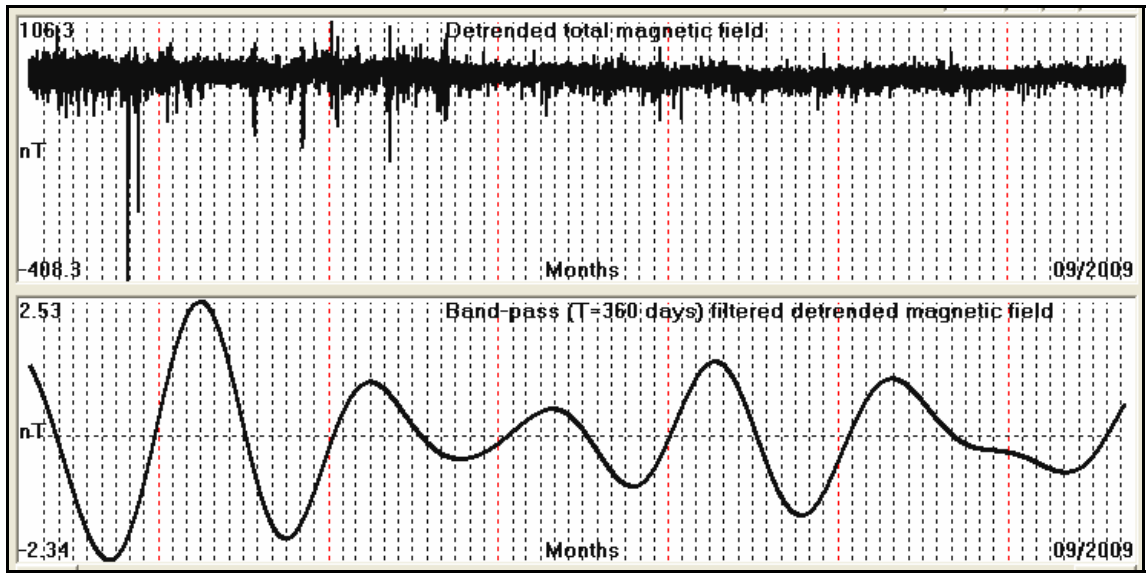

**Fig. 9.** Detrended magnetic raw data (upper graph) and its oscillating component (band-pass filtered with T = 365 days).

**It is evident that the amplitude of the T = 365 days component of the total magnetic field is very small and does not affect macroscopically the upper graph of figure (9). The signal to noise ratio (S/N) in decibels, below noise level (1 day period amplitude adopted as the reference level) of the T = 365 days component of the total magnetic field, has been calculated as S/N = -12.6 db.**

**2.2. Tidal wave analysis.**

**The tidal variation for the study period was calculated by the use of Rudman's method (Rudman et al. 1977). A sample of thee days of tidal variation is presented in figure (10). The actual tidal data have been calculated with a sample interval of one (1) minute which results to 1440 values per a day. The entire study period tidal data are presented in figure (11).**

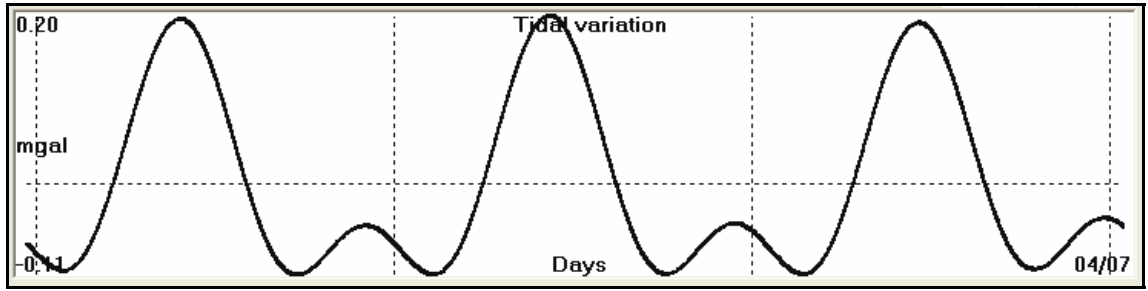

Fig. 10. Three day's sample of the tidal variation. The vertical scale is expressed in mgals since the related computer code was addressing the problem of tidal corrections (vertical component) in gravity surveys.

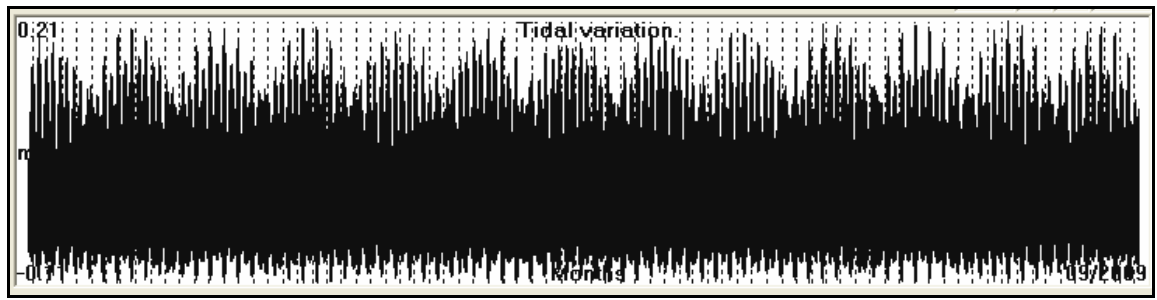

Fig. 11. Tidal variation for the entire study period (20030523 – 20090907).

The application of spectral analysis in the data of figure (11) revealed the presence of the well-known tidal components in the range of T = 30 days, 14 days, 9 days, 24 hours, 12 hours and 8 hours. The dominant components in this data set are the one of 24 hours and its second harmonic of 12 hours. The latter is shown in the following figure (12).

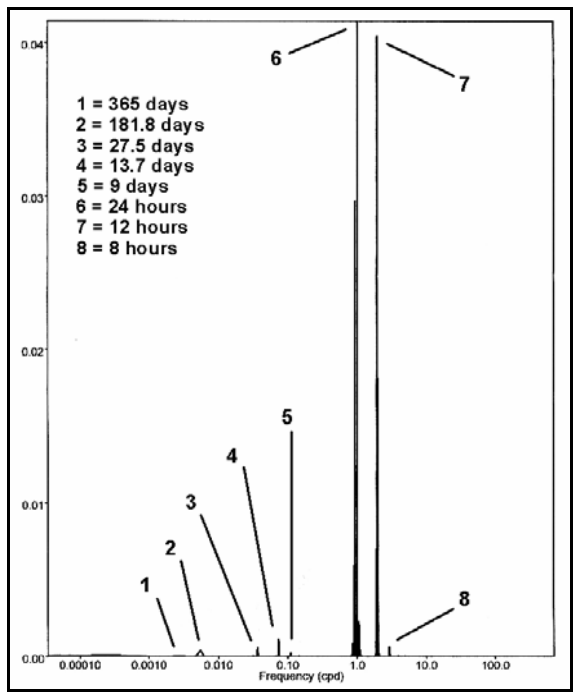

Fig. 12. Results of spectral analysis of the tidal data of the study period (20030523 – 20090907) are shown. The inside table indicates the periods for the identified tidal components.

The larger period and of lower amplitude components are masked from the presence of the dominant shorter period components. Moreover, the short period ones cannot be related to the T = 365 days oscillating components of the Earth's electric field. Therefore, a low-pass filtering procedure is applied on the tidal raw data aiming into eliminate components with T<24 hours. This is simply performed by re-sampling the original tidal data set at a days sampling interval. Consequently, components with T<24 hours are eliminated (Bath 1974, Kulhanek1976, Claerbout 1976). The result of this operation is shown in figure (13).

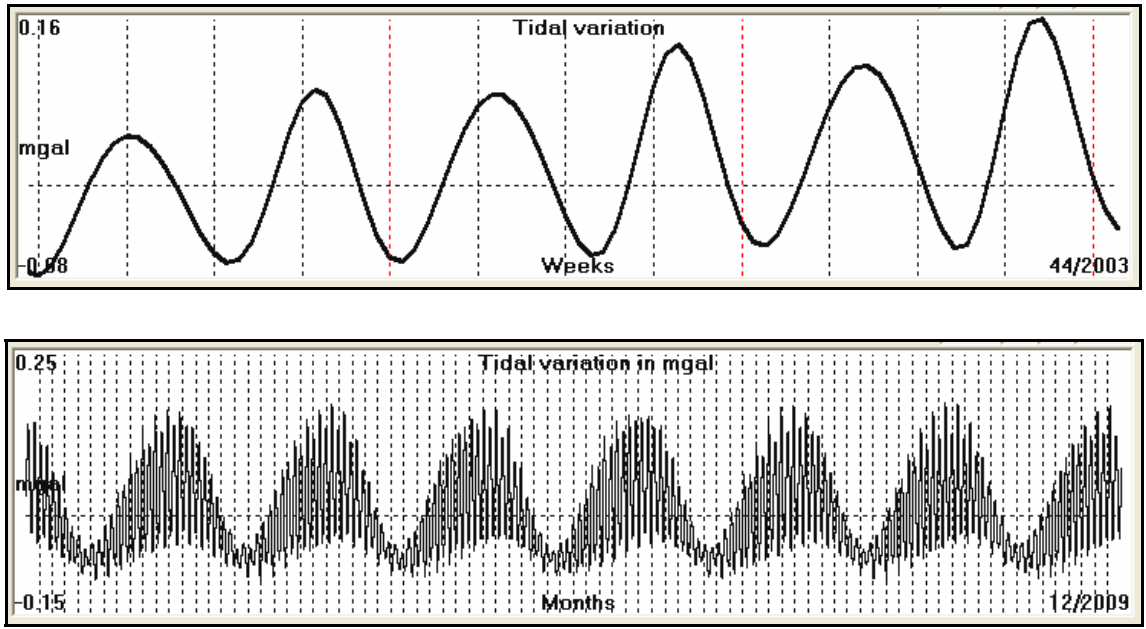

**Fig. 13.** Low-Pass filtered Tidal variation for T<24hours rejection. Three months sample (upper graph) and entire data set (lower graph) are presented.

The data set of figure (13) has been converted from time domain to frequency domain and the corresponding results are presented in figure (14). The inside table indicates the typical tidal basic components and their related higher harmonics.

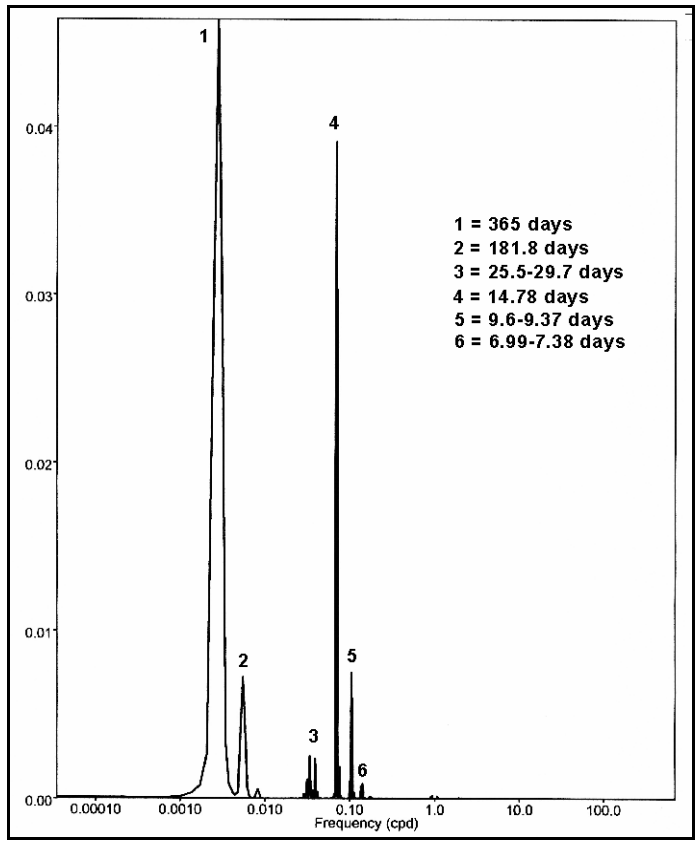

1 = 365 days
2 = 181.8 days
3 = 25.5-29.7 days
4 = 14.78 days
5 = 9.6-9.37 days
6 = 6.99-7.38 days

**Fig. 14.** Spectral analysis of the tidal data set of figure (13). Inside table indicates the basic tidal components and their higher harmonics.

What is clearly shown in figure (14) is the presence of the tidal components of T = 365 days and its higher harmonics while the same is valid for the no. (3) and no. (4) components. The no. 4 component plays an important role in triggering the occurrence of large EQs (Thanassoulas, 2007).

## 3. Discussion - Conclusions

What is made clear up to this point is the fact that the recorded Earth's electric field, the Earth's magnetic field and the tidal waves present some common features in terms of their frequency content. The question that rises is: is there any strong physical connection that relates each one with the others?

Firstly we will examine the case of the Earth's magnetic field variations by hypothesizing it as being the cause of the generation of the observed Earth's electric field oscillation. In the following figure (15) are compared the Earth's magnetic field (lower graph) to the same period Earth's oscillating electric field (upper graph).

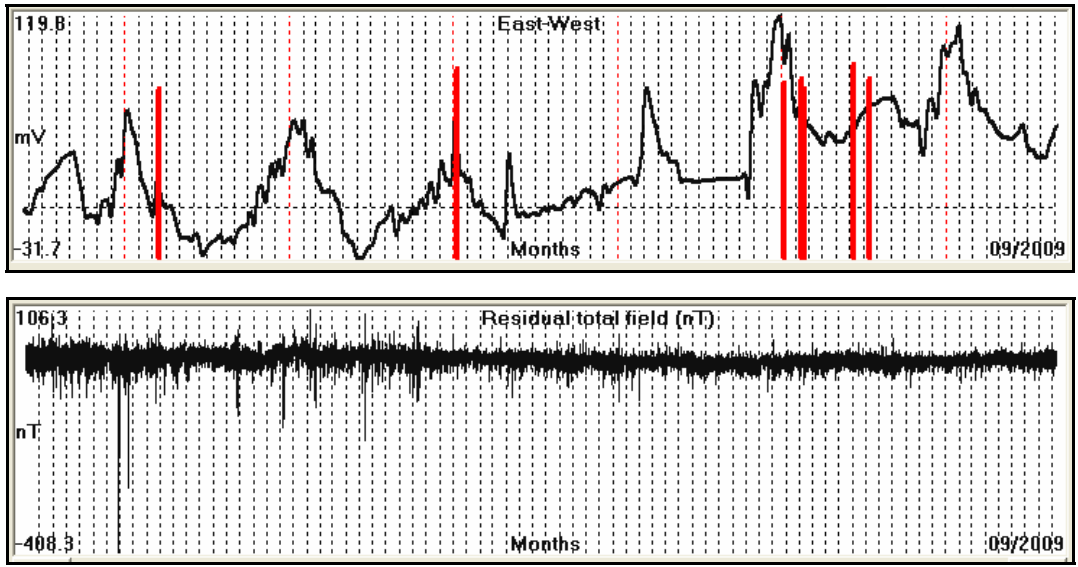

**Fig. 15.** Comparison of the Earth's oscillating electric field (upper graph) to the same period corresponding magnetic field (lower graph).

From figure (15) it is evident that the two graphs present a quite different character. The magnetic field shows an oscillatory (T = 365 days) component which has been calculated to -12.6 decibel bellow the 1 day period noise while the Earth's oscillating field shows a neat similar oscillation which stands +22.3 decibels above the same 1 day period noise. Consequently, the magnetic field cannot produce the observed oscillation (T = 365 days) in the Earth's electric field since it is at - 34.9 decibels bellow the reference noise level. Therefore, the interaction of the Earth's magnetic field with the Earth's ground as a generating mechanism of the observed Earth's electric field oscillation must be rejected.

Next, we will examine the effect of the tidal waves upon the Earth and in particular the tidal wave which exhibits a yearly period. The latter is generated by the elliptic orbit of the Earth around the Sun. This is demonstrated in the following figure (16). In this figure (left) have been marked the characteristic astronomical dates (Solstices, Equinoxes, Aphelion, Perihelion) but of particular interest are the Perihelion and Aphelion. At the perihelion, the Earth is at its closest distance from the Sun while at the Aphelion is at its largest distance from it. Therefore, the Sun applies on the Earth an oscillating attraction with a yearly period. Garland (1971), has pictorially demonstrated the Earth's shape deformation due to this attraction which is shown in figure (16) right. In this drawing what is highlighted is the effect of Earth's shape deformation in combination with its daily rotation. The latter generates the observed daily tidal gravity variations.

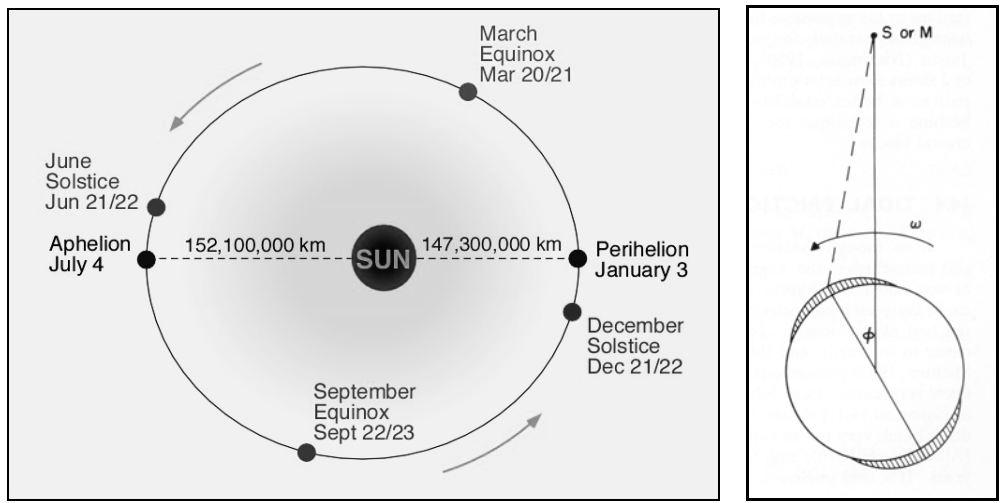

**Fig. 16.** Earth's orbit around the Sun along with specific astronomical dates (left) and its shape deformation due to its daily rotation (Garland, 1971).

In a similar way the Sun's attraction upon the Earth generates a yearly oscillating shape deformation. This is demonstrated in the following figure (17).

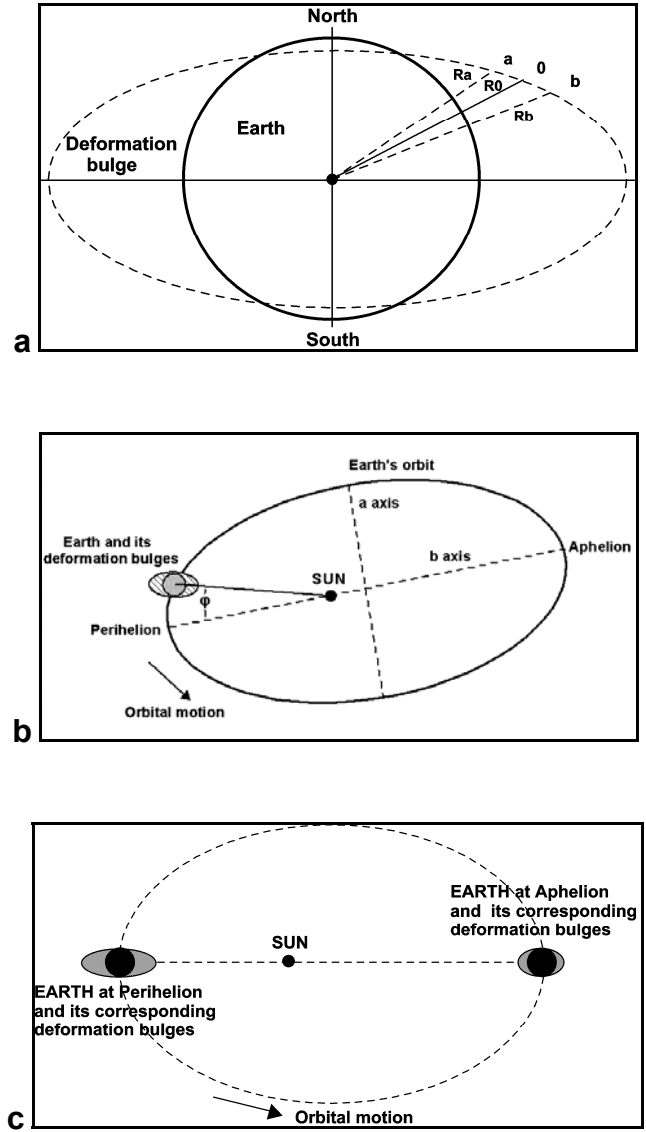

Fig. 17. a = Earth's shape deformation due to Sun's attraction. The radius (Ra, R0, Rb) of the Earth changes accordingly. b = Earth's shape deformation at a place of distance φ from the perihelion, c = Earth's shape deformation at perihelion and aphelion.

What is immediately clear from figure (17) is that the oscillating deformation of the Earth's shape presents its maximum value at the perihelion while its lowest is at aphelion. Consequently, the lithosphere will suffer a similar deformation. From figure (17a) it is evident that the deformation of the Earth's shape will be locally larger in an E-W direction (constant larger Earth radius at E – W direction at a specific latitude) than the N – S one (from the decrease of the Earth's radius towards north).

Thanassoulas (2007) had suggested that the combination of a long term (of some years) stress increase of a seismogenic area combined to the oscillating (short wave lengths) stress component due to tidal waves resulted, at a quite large percentage, into triggering large EQs at the peaks of the tidal waves. This model of triggering large EQs will be tested in this case. Following this model and the classical seismology theories that an EQ occurs at a critical stress level it is expected that large EQs which occurred during the study period must have taken place at the maxima and minima of the yearly lithospheric oscillation, in other words at the Perihelion and Aphelion.

In the following figure (18) the dates of Perihelion (red bars) and Aphelion (green bars) are compared to the large EQs of Table – 1. The numbering of the EQs is the same as in Table – 1.

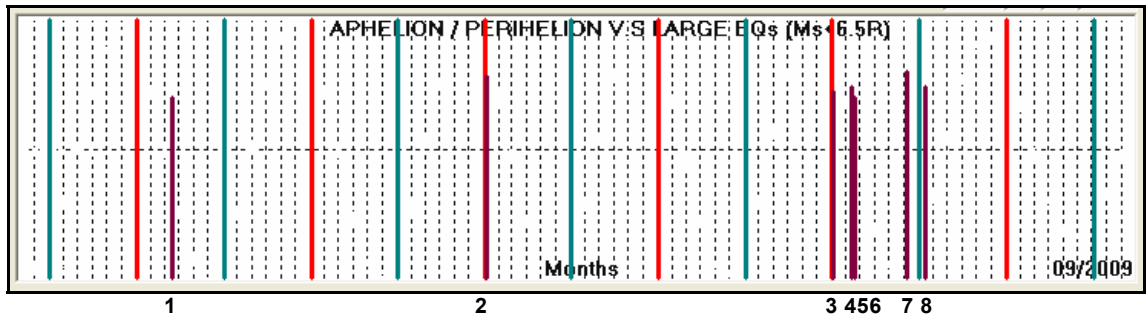

Fig. 18. Comparison of dates of occurrence of large EQs (brown bars, numbers of table - 1) to the Perihelion (red bars) and Aphelion (green bars).

It is very interesting that only one EQ (no.1) is not related to any Perihelion or Aphelion date. The no. (2, 3) EQs coincide very closely to the same short period Perihelion while the no. (4, 5, 6) are clustered closely to the corresponding Perihelion and no. (7, 8) are clustered closely to the corresponding Aphelion. In terms of statistics figure (18) provides the following results:

Failed : (1) = 12.5%

Exact date : (2, 3) = 25.0%

Clustered : (4, 5, 6, 7, 8) = 62.5%

If we consider that "close clustering" of the EQs, either at Perihelion or Aphelion, is a satisfactory result then, by ignoring the no. (1) EQ which failed, the statistical result that favors the postulated mechanism rises to 87.5% which is considered as a quite acceptable value.

It was mentioned earlier that the oscillating deformation of the Earth's shape gives rise to a similar character corresponding deformation of the lithosphere. Consequently, the lithospheric mechanical oscillation causes the generation of an oscillating electric field through various physical mechanisms.

One of the most recently presented physical mechanisms involves the generation of p-hole charge carriers during asymmetric rock stress, and the resulting underground currents and ULF magnetic fields (Freund, 2006, 2007a, b, c). In the case of the present study, asymmetric lithospheric stress is caused by the related asymmetric deformation of the lithosphere (see figure 17a). The latter mechanism refers to the crystalline structure grid deformation of a specific rock formation which is stress loaded nearly to fracture level and therefore takes place at microscopic level. No matter what is the actual mechanism that generates the charge carriers, in a macroscopic level the entire phenomenon behaves more or less as a large scale piezoelectric one. The latter has been verified in many cases of large EQs when the anomalous observed pre-earthquake electric field follows more or less the laws of piezoelectricity, due to the large content of quartzite in the crust (Clark et al. 1924, Taylor 1964), as far as it concerns the generated electric field as a function of applied stress and the occurrence of the large EQ at the characteristic saturated potential level, when the strain of the rock (seismogenic area is no longer linear vs. time increase (Thanassoulas, 2007).

If the latter mechanism is valid, then, the generated electric field will resemble the shape deformation mode of the Earth. That is, the electric field will present peak values at the times when the Earth's deformation is maximum or minimum. In the next figure (19) a comparison is made of the observed oscillating Earth's electric field to the peaks of the Earth's deformation (perihelion, aphelion).

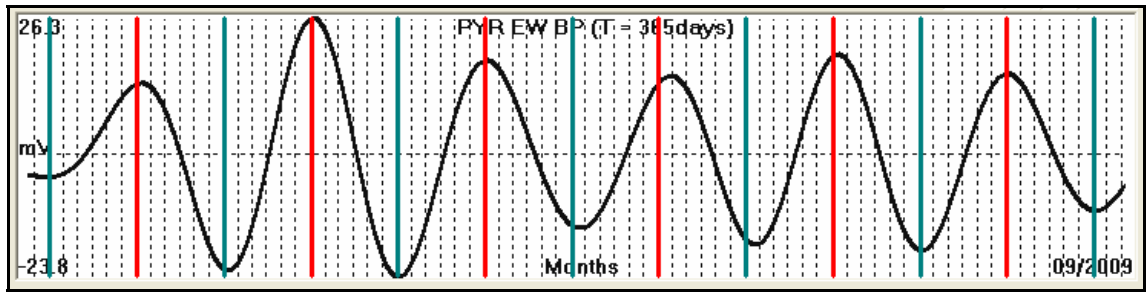

Fig. 19. Comparison of the Earth's oscillating electric field to the same period of time Perihelion (red bars) and Aphelion (green bars).

Figure (19) shows what was expected from the postulated physical mechanism. Some small discrepancies observed of the order of a few days could be attributed to intense short wave preexisting noise in the raw data.

Next we will compare the tidal component of T = 365 days to the dates of perihelion and aphelion. The latter is shown in the following figure (20).

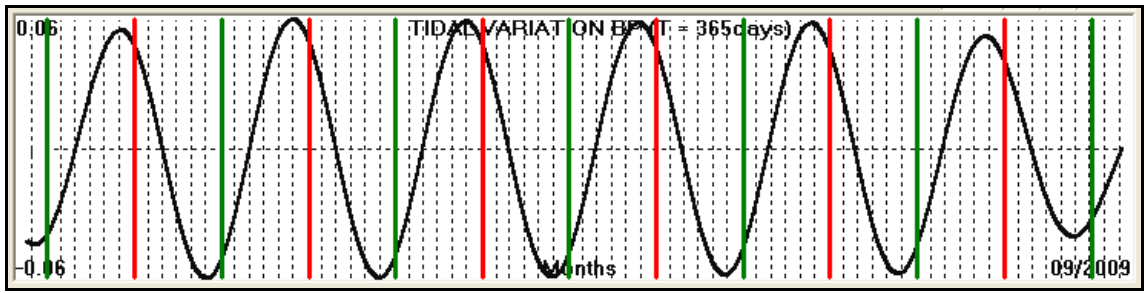

**Fig. 20.** Comparison of tidal component (black line) of T = 365 days to the dates of perihelion (red bars) and aphelion (green bars). The vertical scale is in mgals.

It is made clear from figure (20) that there is a lag between the peaks of the tidal component and the dates of perihelion and aphelion. A close inspection of the graph indicates that this time lag is at average of the order of one month. This lag can be explained by the same mechanism for generating tidal bulges on the Earth (Stacey, 1969), just by substituting the Moon tidal effect by the one generated by the Sun. This lag could be quite well attributed to dissipative processes in the tidal response.

So far, the observed Earth's oscillating electric field complies to the postulated Earth's shape deformation model and with the tidal components of T = 365 days. What is left for further test is the magnetic field (the T = 365 days component) in relation to the perihelion and aphelion dates. This comparison is shown in the following figure (21).

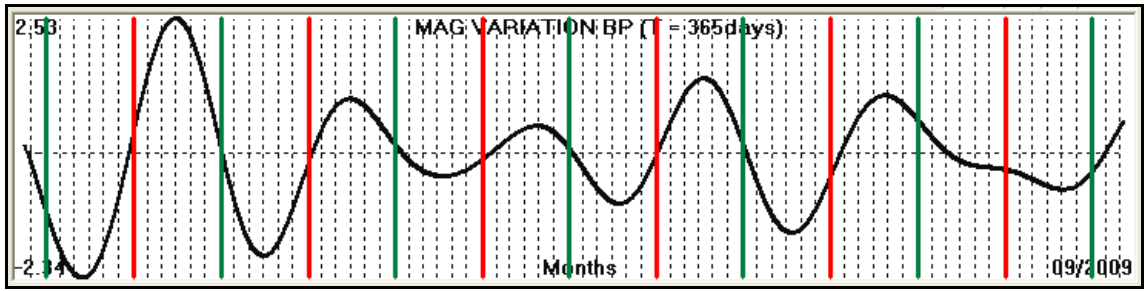

**Fig. 21.** Comparison of the magnetic component of T = 365 days with the perihelion (red bars) and aphelion (green bars). The vertical scale is in nT.

It is evident from figure (21) that the magnetic component of T = 365 days is not correlated directly to the perihelion or aphelion dates. Actually it presents a time lag of three months which could be categorized as "no correlation at all" or there could be (?) a physical mechanism which justifies such a phase shift. The fact that the perihelion and aphelion dates coincide with the "zero-cross" values of the magnetic field component is tempting for a further analysis in the future.

The results already presented in figures (18, 19, 20, 21) are interrelated in the following figures (22, 23, 24). The Earth's electric field (T = 365 days) is presented along the corresponding tidal variation and the same period large EQs and perihelion - aphelion dates in figure (22).

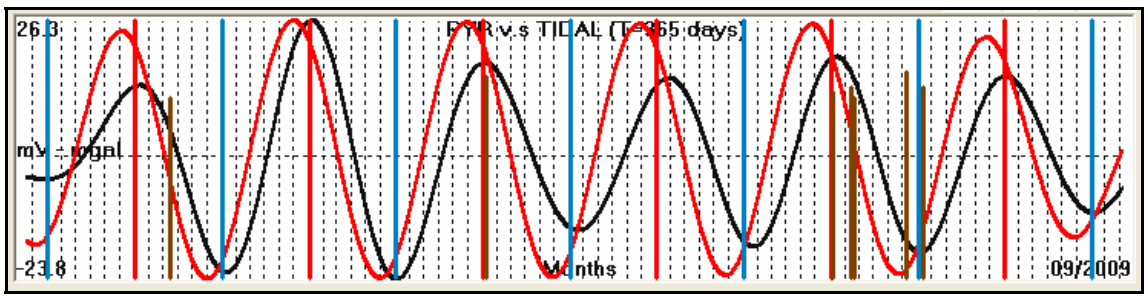

**Fig. 22.** Earth's electric field of T = 365 days (black line) vs. the same period tidal variation (red line), perihelion (red bars) – aphelion (green bars) dates and the same period large EQs (brown bars).

The strong correlation (with one month's lag) of the tidal variation with the Earth's electric field, the perihelion / aphelion dates and the time of occurrence of the large EQs is more than evident.

In figure (23) a simultaneous presentation is made of the Earth's electric field vs. the total Earth's magnetic field, the perihelion – aphelion dates and the time of occurrence of the large EQs.

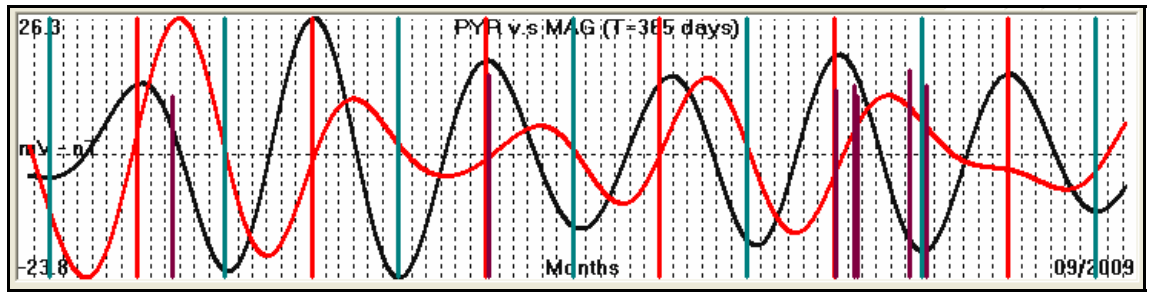

**Fig. 23.** Earth's electric field of T = 365 days (black line) vs. Earth's same period total (red line) magnetic field, perihelion (red bars) – aphelion (green bars) dates and the same period large EQs (brown bars).

A closer observation of figure (23) shows that the Earth's electric field is not related to the observed Earth's magnetic field, as it would be expected by the application of Maxwell's laws. For example, during the first two years of the recording what is observed is a clear increase of the electric field amplitude although the magnetic field decreases at the same period. The same behavior of the electric field is observed during the last two years of the recording. Therefore, it is concluded that the recorded Earth's electric field is not generated by any inductive procedures of the magnetic field on the Earth's ground.

Finally, a comparison is made of the tidal variation with the corresponding Earth's magnetic field, the perihelion – aphelion dates and the time of occurrence of the large EQs. The latter is presented in figure (24).

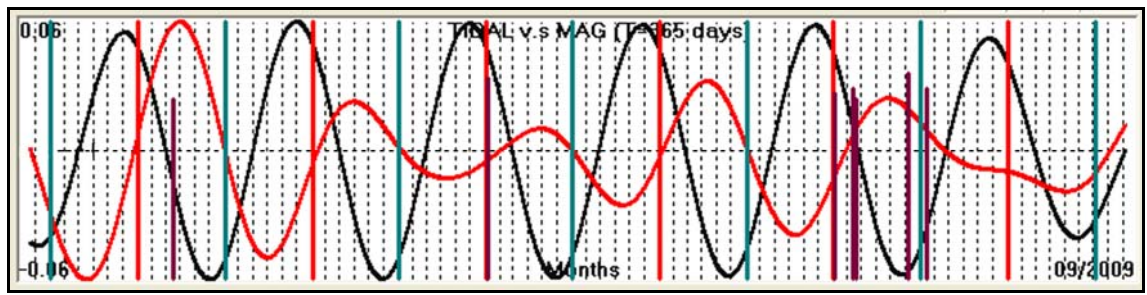

**Fig. 24.** Tidal variation of T = 365 days (black line) vs. Earth's same period total (red line) magnetic field, perihelion (red bars) – aphelion (green bars) dates and the same period large EQs (brown bars).

The observed change in amplitude of the magnetic field vs. the almost stable in amplitude tidal variation suggests that there is no strong physical relation between the tide generation mechanism (Earth's orbiting around the sun) and the oscillating magnetic field of the Earth.

After summarizing all the previous results it is concluded that:

- The Earth's orbit around the Sun generates an oscillating (T = 365 days) deformation of the Earth's shape and its lithosphere accordingly.

- The lithospheric deformation, observed at a certain place on the Earth's surface, varies at N-S and E-W directions, the E-W direction being the largest one.

- The yearly tidal variation precedes the perihelion – aphelion dates for a month probably due to tidal friction mechanism.

- When the generated stress, by the latter deformation, is combined to the stress increase which is developed prior to a large EQ at a seismogenic area due to tectonic causes, then, at critical stress level, an increasing in amplitude oscillating (T = 365 days) electric field is generated due to locally triggered large scale piezoelectric phenomena.

- The generated oscillating electric field is not related to or induced by the Earth's magnetic field.

- Close to the time of occurrence of the large EQ the observed oscillating electric field oscillates in phase for both components of the N-S and E-W directions.

- The triggered large EQs take place very close to the perihelion – aphelion times when the lithospheric deformation is at its peak (maximum – minimum) value.

- **It seems that long-term monitoring of the Earth's electric field can provide indicators whether a regional seismogenic area has enter a critically stress load period thus suggesting a medium- long term large EQs prognosis methodology.**